\documentclass[twocolumn,showpacs,preprintnumbers,amsmath,amssymb,prl]{revtex4}

\usepackage{epsfig}
\usepackage{dcolumn}
\usepackage{bm}
\usepackage{latexsym}

\begin{document}
\preprint{APS/123-QED}

\title{Nonequilibrium nuclear-electron spin dynamics in semiconductor
quantum dots}

\author{D. H. Feng, I. A. Akimov, and F. Henneberger} \affiliation{
Institut f\"ur Physik, Humboldt Universit\"at zu Berlin,
Newtonstr.15, 12489 Berlin, Germany}

\date{\today}

\begin{abstract}
We study the spin dynamics in charged quantum dots in the situation
where the resident electron is coupled to only about 200 nuclear
spins and where the electron spin splitting induced by the
Overhauser field does not exceed markedly the spectral broadening.
The formation of a dynamical nuclear polarization as well as its
subsequent decay by the dipole-dipole interaction is directly
resolved in time. Because not limited by intrinsic nonlinearities,
almost complete nuclear polarization is achieved, even at elevated
temperatures. The data suggest a nonequilibrium mode of nuclear
polarization, distinctly different from the spin temperature concept
exploited on bulk semiconductors.
\end{abstract}



\maketitle

The hyperfine interaction of the electron with the nuclear spins in
semiconductor quantum dots has become recently a focus of both
theoretical and experimental research. In the absence of external
magnetic fields, it governs the timescale on which the electron spin
can be stored and manipulated. The dynamical scenario depends
sensitively on the state of the nuclear system as well as on the
type of average performed in the
experiment\cite{MER02,BML05,KLG0203}. On the other hand, continuous
pumping of the electron spin can generate through hyperfine-mediated
spin flip-flops a dynamical nuclear polarization (DNP). It is
important to distinguish between nonequilibrium and equilibrium DNP
\cite{OO84}. The nonequilibrium DNP is destroyed by the nuclear
dipole-dipole interaction with a time constant  ($\tau_\text{d-d}$)
of about 10$^{-4}$ s. In bulk semiconductors, because of the huge
number of nuclear spins seen by the electron, the formation time
($\tau_\text{F}$) of the DNP is much longer than the dipole-dipole
decay, ruling out a significant nonequilibrium mode. However, spin
temperature cooling in an external magnetic field can produce an
equilibrium DNP. The field Zeeman-splits the nuclear states
sufficiently up and selectively pumping one of the populations via
the flip-flop process creates a nuclear spin temperature different
to that of the lattice \cite{AP53}. The decay of the equilibrium DNP
requires dissipation of the Zeeman energy by spin-lattice
interaction which takes place on timescales of a second and beyond.
It has been argued that an external field is not required for
quantum dots, as the hyperfine Knight field is strong enough to
ensure spin cooling conditions \cite{L06,AFH06}. For InAs/GaAs
structures, the presence of the DNP is directly displayed by a
zero-field splitting of the dot emission \cite{L06}. The Overhauser
field can reach here the 1-Tesla range and gives rise to strong
nonlinearities in the electron spin occupation \cite{B06}. The DNP
formation time estimated from indirect measurements is
$\tau_\text{F} \sim$ 1 s \cite{L06,B06}. In this Letter, we
investigate Stranski-Krastanov CdSe/ZnSe quantum dots which have a
2-3 orders of magnitude smaller volume size. We directly demonstrate
that these quantum dots realize the unique situation $\tau_\text{F}
< \tau_\text{d-d}$ allowing for - unlike the vast amount of previous
work on semiconductors - the formation of a strong nonequilibrium
DNP.

\begin{figure}
\centering
\includegraphics[width = 8.0 cm]{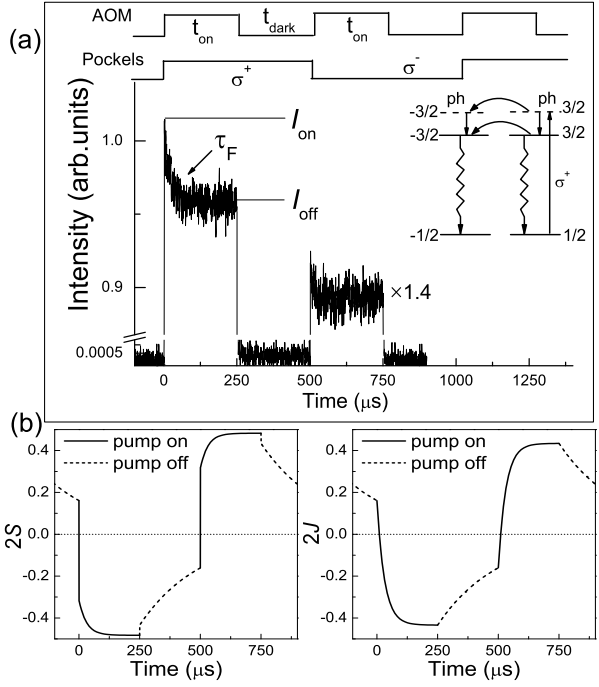}
\caption{Electron-nuclear dynamics as revealed by quasi-resonant
excitation of the trion feature. (a) Optical excitation trains in
alternating polarization mode (top) as well as PL transients (main
panel) recorded at $T$ = 60 K and $B=0$. Excitation intensity is 2
kW/cm$^2$. Excitation and detection (1-LO-phonon below) is on the
low-energy wing of the ensemble PL band centered at about 2.43 eV.
Detection is in $\sigma^+$ polarization. The different signal levels
in $\sigma^+$ and $\sigma^-$ pumping reflect the degree of hole spin
relaxation in the trion \cite{FHA03}. $\sigma^-$ detection just
reverses these levels. For more details about the experimental setup
see \cite{AFH06}. Right inset: Optical transition scheme underlying
the electron spin pumping for $\sigma^+$ photons. A flip of the 3/2
hole spin in the trion directs the excitation to the other arm so
that a spin-down electron is left behind after recombination (ph:
trion-LO-phonon state). (b) Calculated electron and nuclear spin
transients from equations (\ref{equations}) with parameters $N =
200, \tau_\text{e} = 50$ ns, $\hbar A = -6~\mu$eV, $\tau_\text{c} =
1.2$ ns, $\tau_\text{d-d} = 250~\mu$s.}
\label{fig:Fig1}
\end{figure}

The quantum dot structures are grown by molecular beam epitaxy.
Making use of the native n-type of these materials, dots charged
with a resident electron are formed under appropriate stoichiometry
conditions. Transmission electron microscopy provides a dot height
of below 2 nm and a lateral extension below 10 nm \cite{L02}. The
optical measurements are performed in a confocal arrangement with
the propagation direction of incident and emitted light parallel to
the [001] growth axis. The sample is placed in a He-flow cryostat
capable of variable temperature and magnetic fields up to $B = 5$ T.
Quasi-resonant excitation of the trion feature (2 electrons, 1 hole)
by circularly polarized light constitutes a very efficient spin
pumping mechanism for the resident electron \cite{AFH06}. In turn,
the secondary emission from the trion directly monitors the electron
spin polarization achieved. Trains of rectangular light pulses with
a duration $t_\text{on}$ spaced by a dark time $t_\text{dark}$ are
generated from a \textit{cw} laser by means of an acousto-optical
modulator (Fig. \ref{fig:Fig1}a). Use of a properly triggered
Pockels cell allows for the generation of sequences with alternating
$\sigma^+/\sigma^-$ polarization or with $\sigma^+$ polarization
only. A difficulty of the measurements is an increasing number of
dark counts when $t_\text{dark}$ exceeds considerably $t_\text{on}$.
The data presented below are selectively verified on a single-dot
level, systematic studies at reasonable integration times are made
on ensembles with a much better signal-to-noise ratio. The external
magnetic field is always directed along the electron spin.

\begin{figure}
\centering
\includegraphics[width=6.5 cm]{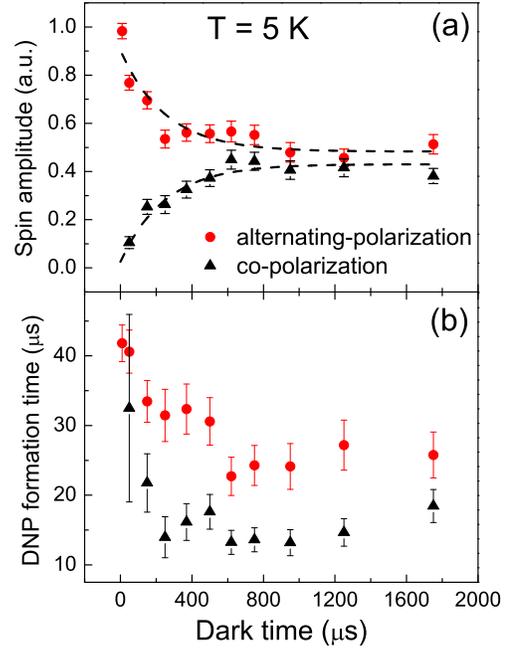}
\caption{Spin amplitude (a) and DNP formation time (b) as functions
of the dark time in both polarization modes ($B = 0$). The dashed
curves are single-exponential fits with a time constant of 250
$\mu$s.}
\label{fig:Fig2}
\end{figure}

Fig. \ref{fig:Fig1} summarizes the main features of the experiment.
Because of the anti-parallel spins of the two electrons, the total
trion spin is defined by the hole and is thus $\pm3/2$. This
provides the circular polarization selection rules
$|\pm1/2\rangle+\sigma^{\pm} \leftrightarrow |\pm3/2\rangle$ of the
electron-trion transition. In order to suppress stray light, the
trion is quasi-resonantly excited with 1-LO phonon energy in excess
through a polaron-like state. This state has the same spin structure
as the trion ground-state and obeys hence the same selection rules.
Excitation, say with $\sigma^+$ photons, removes thus selectively a
spin-up electron. The polaron-trion state relaxes rapidly to the
trion ground-state which subsequently radiatively decays within
about 500 ps. When the hole spin-flips, a spin-down electron is left
behind. More precisely, denoting by $\varrho_{\downarrow\downarrow}$
and $\varrho_{\uparrow\uparrow}$ the diagonal elements of the
electron spin density matrix, $\varrho_{\downarrow\downarrow}$ is
increased at the expense of $\varrho_{\uparrow\uparrow}$ or, in
other words, an average spin  $S =
(\varrho_{\uparrow\uparrow}-\varrho_{\downarrow\downarrow})/2 <0$
just opposite to the excitation photon momentum is formed. The
associated depletion of the spin-up initial state in absorption
decreases the trion excitation rate $\propto
\varrho_{\uparrow\uparrow} = 1/2+S$ and, in the same way, the
photoluminescence (PL) signal reemitted by the dot (independent on
the photon polarization detected). The experimental PL transients in
Fig. \ref{fig:Fig1} reveal therefore directly an increasing electron
spin alignment during pumping. The spin amplitude $\Delta S$
generated is proportional to $
(I_\text{on}-I_\text{off})/I_\text{off}$ and the characteristic time
for its formation ($\tau_\text{F}$) is taken at the 1/e point of the
transients. Fig. \ref{fig:Fig2} depicts both quantities as a
function of the dark time. During the dark period, electron as well
as nuclear spin decay down to a certain level. Restarting pumping
enables us to study the spin dynamics at different initial
polarizations. Complete loss of the spin memory from the previous
pulse has happened when the amplitudes in alternating and
co-polarization become equal. As evident from Fig. \ref{fig:Fig2},
this time is considerably longer than the few 10 $\mu$s needed for
the build-up of the amplitude.

The electron-nuclear spin dynamics revealed by the above
measurements is well described by the equations
\begin{eqnarray}\label{equations}
\frac {\text{d}S}{\text{d}t} &=& \pm \frac{1}{2} P-\frac {S}{\tau
_\text{e}} + N \frac {J-S}{\tau_\text{hf}},\\ \nonumber
\frac {\text{d}J}{\text{d}t} &=&  \frac {S-J}{\tau_\text{hf}} -
\frac {J}{\tau_\text{d-d}},
\end{eqnarray}
considering neither inhomogeneity of the hyperfine coupling nor
collective effects in the nuclear system \cite{CCG07}. The $\pm$
sign at the spin pumping rate $P$ stands for $\sigma^\mp$ excitation
and $N = a N_\text{L}$ with the effective number $N_\text{L}$ of
lattice atoms surrounding the electron and the isotope abundance
$a$. The isospin of Cd and Se is 1/2 so that the nuclear average
spin $J$ per isotope is analogously defined as for the electron.
$\tau _\text{e}$ summarizes the lifetime of the electron in the dot
as well as all spin relaxation processes aside from the hyperfine
interaction. If only the hyperfine contribution (characterized by a
single time $\tau_\text{hf}$) is accounted for, these equations
ensure total spin conservation $\dot{S}+ N \dot{J}=0$. The linear
system is easily solved analytically yielding two time constants by
which the steady-state values $S_\text{st} = \pm P
\tau_\text{e}(\tau_\text{hf}+\tau_\text{d-d})/2(\tau_\text{hf}+\tau_\text{d-d}+N
\tau_\text{e})$ and
$J_\text{st}=S_\text{st}\tau_\text{d-d}/(\tau_\text{hf}+\tau_\text{d-d})$
are approached. The sole electron spin pumping scenario is studied
by eliminating the hyperfine coupling by an external magnetic field
of $B$ = 100 mT \cite{AFH06}. The electron lifetime is
power-dependent $1/\tau_\text{e}= P+\eta P^\gamma$, both
intrinsically due to the excitation process itself as well as due to
dot recharging by carriers generated in the environment as a side
effect ($\eta\sim 1, \gamma \lesssim 1$). Specifically, this
generates a spin polarization $2S = P \tau_\text{e}$ of close to 0.5
depending only weekly on the pumping rate. At the power of the
present measurements $\tau_\text{e} = 50$ ns. During pumping, it
holds thus $\tau_\text{e} \ll \tau_\text{d-d}$ which provides in
lowest order
\begin{eqnarray*}
\delta S(t) &=& \delta S_\text{0} \text{e}^{-t/\tau_1}-\delta J_\text{0} K (\text{e}^{-t/\tau_1}-\text{e}^{-t/\tau_2} ),\\
\delta J(t) &=& \delta J_\text{0} \text{e}^{-t/\tau_2},
\end{eqnarray*}
for the deviations from steady state. The times $\tau_1$ and
$\tau_2$  are very much different for $N \gg 1$. The short component
$\tau_1 = \tau_\text{e}\tau_\text{hf}/(\tau_\text{hf}+N
\tau_\text{e})$, not resolved on the timescale of Fig.
\ref{fig:Fig1}, describes the direct response of the electron on
spin pumping, whereas $\tau_2 = (\tau_\text{hf}+ N
\tau_\text{e})/(1+\tau_\text{hf}/\tau_\text{d-d}+N
\tau_\text{e}/\tau_\text{d-d})$ represents the DNP formation time
$\tau_\text{F}$ (shown in Fig. \ref{fig:Fig2}b) and $K = N
\tau_\text{e}/(\tau_\text{hf}+ N \tau_\text{e})$ measures the degree
by which the presence of the DNP increases the electron spin
polarization. During the dark period ($P, S_\text{st}, J_\text{st} =
0$), $\tau_\text{e}$ is orders of magnitude longer so that now the
single-spin flip-flop defines the shortest timescale
($\tau_\text{hf}/N \ll \tau_\text{hf}, \tau_\text{d-d},
\tau_\text{e}$) providing $\tau_1=\tau_\text{hf}/N$,
$\tau_2=\tau_\text{d-d}$, and $K=1$. The short $\tau_1$ enforces $S
\approx J$, while both polarizations decay slowly by
$\tau_\text{d-d}$. Application of periodic boundary conditions
results in spin amplitudes $\Delta S =\Delta S_{\infty}[1 \pm
\exp(-t_\text{dark}/\tau_\text{d-d})]$ for alternating and
co-polarization mode, respectively, where $\Delta S_{\infty}= -K
J_\text{st}~(t_\text{on} \gg \tau_\text{F}$). The experimental data
in Fig. \ref{fig:Fig2}a follow closely this prediction yielding
$\tau_\text{d-d} = 250~\mu$s. The formation time of the DNP is hence
about one order of magnitude shorter than its decay by the
dipole-dipole interaction.

The linewidth broadening of the electron spin levels is determined
by the lifetime of the off-diagonal density matrix elements
$\varrho_{\uparrow\downarrow}$. Denoting this time by
$\tau_\text{c}$ and assuming $(A/N_\text{L})\tau_\text{c} \ll 1$
leads to the standard expression $1/\tau_\text{hf}=1/T_\text{1e}=
(A/N_\text{L})^2 \tau_\text{c}/(1+
\Delta\omega^2_\text{B}\tau^2_\text{c})$ \cite{DP73} where
$\Delta\omega_\text{B}$ represents the electron Zeeman splitting
\footnote{The splitting of the nuclear levels can be neglected.}.
For no external field, the nuclear Overhauser field associated with
the DNP provides $\Delta\omega_\text{B}= a A J$. This term
introduces nonlinearities in the dynamics which inhibit the
formation of the DNP by increasing the formation time \cite{B06}. A
distinct feature of the experimental findings is a continuous
increase of $S$ during spin pumping, also in alternating
polarization mode (Fig. \ref{fig:Fig1}a). Here, the nuclear spin
reverses sign so that $1/\tau_\text{hf}$ changes from slow to fast
to slow and the electron spin would passes through a minimum if $a A
\tau_\text{c} $ was markedly larger than 1. A weak Overhauser
feedback is also signified by the dark-time dependence of the DNP
formation time (Fig. \ref{fig:Fig2}b). Indeed, $\tau_\text{F}$ is
longer for smaller $t_\text{dark}$ because of a stronger initial
DNP, but the overall change is only about 30 $\%$. Solving equations
(\ref{equations}) numerically with the Overhauser field included
yields $a A \tau_\text{c} = 1.4$. This value is consistent with the
asymmetries of the spin transients observed in a weak external
magnetic field \cite{AFH06}. The present quantum dots define thus a
regime where the DNP-induced spin splitting does not exceed markedly
the broadening of the Zeeman levels.

The fast DNP formation time of a few 10 $\mu$s is in accordance with
the hyperfine parameters of CdSe quantum dots. The electron
wavefunction $\varphi_\text{e}$ obtained by calculating the energy
position of the ensemble PL provides an average number $N_\text{L} =
8 / (v_0\int \text{d}V \varphi_\text{e}^4(r)) \approx 1600$ ($v_0$:
volume of unit cell) \cite{KPR01}. Accounting for only the major
contribution from the $^{111,113}$Cd isotopes ($a = 0.125$), it
follows that merely $N = 200$ atoms carry a nuclear moment. For
$\mu_I^\text{Cd}$ = -0.6, we estimate $\hbar A = -6 \mu$eV
\cite{N79}. The disappearance of the DNP-related spin amplitude in
an external field of some 10 mT ($\Delta \omega_\text{B}=9$ ns$^{-1}
B/100$ mT) provides $\tau_\text{c} \sim 1$ ns \cite{AFH06}.
Combining these data, it follows $a A \tau_\text{c} \sim 1$ and
$\tau_\text{hf}\simeq 25~\mu$s. Hyperfine flip-flop and electron
lifetime ($N \tau_\text{e}\simeq 10~\mu$s) contribute thus about
equally to the DNP formation time ($K \sim 0.5)$. Spin transients
computed for the above parameters predict virtually the same nuclear
and electron spin polarizations (Fig. \ref{fig:Fig1}b). Full closing
of the hyperfine flip-flop rates is experimentally corroborated by
single-dot measurements. The zero-field transients approach the same
final value as the fast transients at $B = 100$ mT of $2S \approx
0.5$.

\begin{figure}
\centering
\includegraphics[width=8.5 cm]{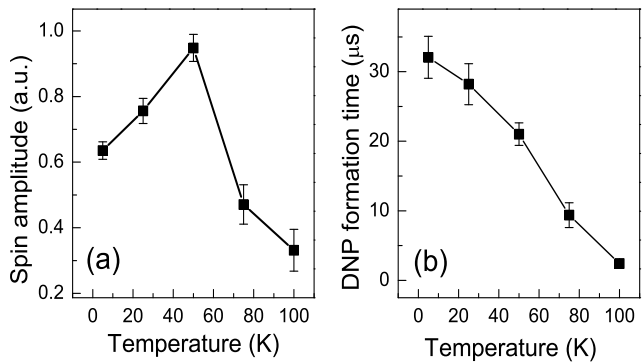}
\caption{Temperature of dependence of (a) spin amplitude and (b)
formation time for $t_\text{dark}=0$ taken in alternating
polarization mode ($B=0$). }
\label{fig:Fig3}
\end{figure}

The DNP formation time shortens monotonically with temperature,
while the spin amplitude initially slightly increases reaching a
maximum at about 50 K before it starts to decline (Fig.
\ref{fig:Fig3}). A shorter $\tau_\text{F}$ at higher temperature is
expected from the shortening of the electron lifetime. The
temperature increase of the amplitude is consistent with the
behavior of the pumping rate. At given optical power, $P$ increases
first as the hole spin flip in the trion state becomes faster
\cite{FHA03}. Subsequently, the shortening of the electron lifetime
takes over. Single-dot measurements where contribution of nontrionic
origin can be excluded confirm these tendencies.

\begin{figure}
\centering
\includegraphics[width=9.0 cm]{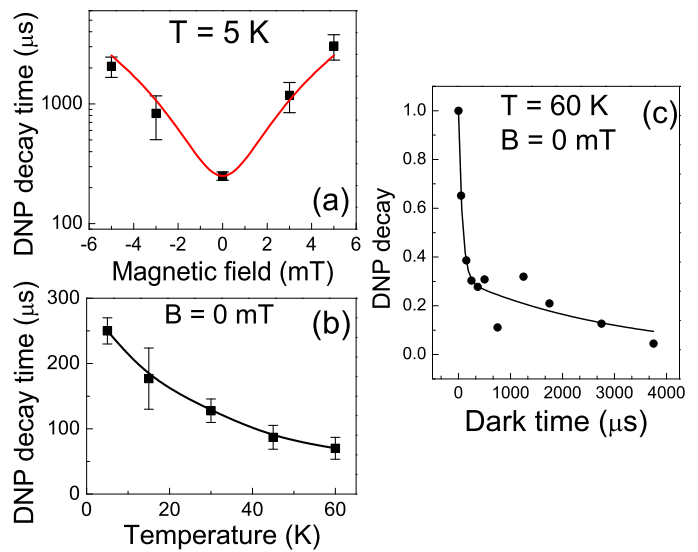}
\caption{Dynamics of the DNP decay. (a) Magnetic field dependence of
the decay time obtained from single-exponential fits to the
experimental $(\Delta S_\text{al} - \Delta S_\text{co})/(\Delta
S_\text{al} + \Delta S_\text{co})$ for lower noise. The line
represents $\tau_\text{d-d}(B)= 250 \mu$s + 91 $\mu$s/mT $B^2$. (b)
Temperature dependence. (c) Decay transient at elevated temperature
on a longer timescale.}
\label{fig:Fig4}
\end{figure}

After elaboration of the formation dynamics we deal finally with the
decay of the DNP by the dipole-dipole interaction.  The $250~\mu$s
decay time is in accord with the interaction energy of
10$^{-5}~\mu$eV of neighboring nuclei. Standard spin cooling is
described by a depolarization time $\propto
B^2/\tilde{B}_\text{L}^2$ where $\tilde{B}_\text{L}$ is the
effective local dipole field \cite{GEK01}. A substantial Knight
field $B_\text{e} \propto S$ added to $B$ \cite{L06,AFH06} would
show up by an asymmetry of the decay transients with respect to the
directions of electron spin and external field. No such asymmetry is
found beyond the experimental resolution. The magnetic-field
dependence of the dipole-dipole time is depicted in Fig.
\ref{fig:Fig4}a. It increases about one order of magnitude for
modest field strength ($B =5$ mT), clearly demonstrating the opening
of a nuclear spin splitting suppressing dipole-dipole-mediated
transitions between different nuclei. The zero-field time decreases
only smoothly with temperature (Fig. \ref{fig:Fig4}b) so that the
relation $\tau_\text{F} < \tau_\text{d-d}$ is maintained up to 100
K. Measurements on an extended timescale (Fig. \ref{fig:Fig4}c)
uncover in addition to the 100 $\mu$s component a second, clearly
smaller part which persists up to the millisecond range. This part
becomes increasingly visible at higher temperature and makes up a
0.2 portion of the total amplitude at 60 K. The origin of the long
persisting DNP needs further investigations. It might be due to the
Se isotopes with lower abundance or spin diffusion between different
dots.

In summary, we have demonstrated the formation of a strong
nonequilibrium DNP at zero external magnetic field. A number of only
a few 100 nuclei and a weak Overhauser field allow for the creation
of the DNP in a time much shorter than in the standard  spin cooling
protocol. The Knight field does not play an essential role. We have
found an intrinsic zero-field dipole-dipole decay time of about 250
$\mu$s. The nuclear polarization is complete in the sense that it
reaches the level of the available electron spin. Minimizing dot
recharging under injection by structure improvement will permit to
approach the ultimate limit $2S, 2J \sim 1$. Signatures of a DNP
formation are seen up to about 100 K. At those temperatures, the
optical spin pumping and read-out via the trion features break down.
Whether or not a DNP can be established at still higher temperatures
can not be decided by the present experiments. The measurements
suggest a very short electron spin correlation time $\tau_\text{c}
\sim 1$ ns under optical pumping. Spin exchange processes with
carriers excited in the dot environment \cite{P82} seem to play an
essential role in Stranski-Krastanov structures.

This work was supported by the Deutsche Forschungsgemeinschaft
within Project No. He 1939/18-1.

\end{document}